\documentstyle[twocolumn,prl,aps]{revtex}

\begin{document}
\draft
\title{Ballistic electron transport through magnetic domain walls}

\author{Jeroen B.A.N. van Hoof,$^{1,2}$~\cite{Sara} 
Kees M. Schep,$^{1,2}$~\cite{Philips} 
Arne Brataas,$^{1,2}$ 
Gerrit E.W. Bauer,$^{1}$ and
Paul J. Kelly$^{2,3}$~\cite{Twente}}
\address{$^1$Laboratory of Applied Physics and Delft Institute of
Microelectronics and Submicrontechnology, Delft University of Technology, Lorentzweg 1, 2628 CJ
Delft, The Netherlands\\
$^2$Philips Research Laboratories, Prof.\ Holstlaan 4, 5656 AA Eindhoven, The Netherlands\\
$^3$ Faculty of Applied Physics, Twente University, P.O.\ Box 217, 5600 AE
Enschede, The Netherlands}

\date{\today}

\maketitle
\begin{abstract}
Electron transport limited by the rotating exchange-potential of domain
walls is calculated in the ballistic limit for the itinerant ferromagnets 
Fe, Co, and Ni. When realistic band structures are used, the
domain wall magnetoresistance is enhanced by orders of magnitude compared to
the results for previously studied two-band models. Increasing the
pitch of a domain wall by confinement in a nano-structured point
contact is predicted to give rise to a strongly enhanced magnetoresistance.
\end{abstract}


The application potential of magnetoresistive effects has rekindled
interest in the study of electrical transport in metallic (Stoner) 
ferromagnets such as Fe, Co, and Ni.
One complicating factor which is still an open problem is the influence
on the transport properties of the magnetic domain structure. 
Domain walls (DWs) result from minimizing the sum of the magnetostatic,
magnetic anisotropy, and exchange energies and they can be driven out of
a material by applying a magnetic field. 
This modifies the transport properties but the magnitude and even the
sign of the magnetoresistance
${\rm MR} \equiv [R_{\rm sat}-R_0]/R_0=[G_0-G_{\rm sat}]/G_{\rm sat}$ 
(where $R_0=1/G_0$ and $R_{\rm sat}=1/G_{\rm sat}$ are the zero-field
and saturation-field resistances, respectively) remain a matter of controversy.

Early experimental studies on very pure iron samples showed a large
MR of up to 90\% at low temperatures\cite{whisker}
which was attributed to percolation
through numerous domains \cite{Berger}. Using a free electron model, 
Cabrera and Falicov\cite{Cabrera} interpreted transport through a single DW as a
tunneling process and the corresponding MR was found to be exponentially small. 
More recently, Tatara and Fukuyama\cite{Tatara2} calculated the DW
conductivity in the clean limit where the mean free path $\ell$ 
resulting from defect scattering is much larger than the wall width
$\lambda_{DW}$.
For a free electron model in a semiclassical approximation they found
an MR which scales like $-n_{DW}/(\lambda_{DW}\overline{k}_F^2)$, where
$n_{DW}$ is the domain wall density and 
$E_F \equiv \hbar ^2 \overline{k}_F^2 / 2m$ is the Fermi energy. This MR 
is also very small for DW widths and Fermi energies of transition metals.
In room temperature measurements of
transport through Ni and Co films exhibiting stripe domain structures,
Gregg {\em et al.}\cite{Gregg} measured significant negative MRs, much
larger than predicted by any of the above theoretical work (but in a regime
where $\ell \leq \lambda_{DW}$). Levy and Zhang\cite{Levy} subsequently
pointed out that spin-dependent impurity scattering can strongly
enhance the (negative) DW-MR. Breaking of the weak localization quantum
correction by the exchange field leads to a positive MR at low
temperature~\cite{Tatara2}. Otani {\em et al.}~\cite{Otani} and
R\"{u}diger {\em et al.}
\cite{Rudiger} measured a positive DW-MR in thin magnetic wires, but up to high
temperatures. 

In the spirit of previous work on the giant magnetoresistance of
magnetic multilayers\cite{Schep1} we study DW scattering in the
ballistic limit, {\em i.e.} in the limit where the defect scattering
mean free path $\ell$ is sufficiently larger than the system size
\cite{Mathon}. 
These results are appropriate for clean point contacts with diameters
$d$ sufficiently smaller than $\ell$. We disregard lateral quantization, assuming
$d \gg \lambda_F$, with $\lambda_F$ the Fermi wavelength.
For perpendicular transport in multilayers, the ballistic MR is of the same
order of
magnitude as the MR in diffuse systems \cite{Schep1,Bauer}.
When $\ell \gg \lambda_{DW}$,
our calculated transport coefficients can be used as boundary
conditions in the
semiclassical Boltzmann equations \cite{Schep2}.

The conductance $G$ is given by
Landauer's formula as:
\begin{equation}
G = \frac{e^2}{h} \sum_{\vec{k}_\parallel} \sum_{\nu\mu}
\left | t_{\nu\mu}({\vec{k}_\parallel}) \right | ^2 ,
\label{eq:land2}
\end{equation}
where $\vec{k}_\parallel$ is the conserved Bloch vector parallel to the DW.
The transmission amplitude of an incoming state $\vec{k}_\parallel \mu$ to an
outgoing state $\vec{k}_\parallel \nu$ through the DW sandwiched by the two
bounding domains of the ferromagnet is denoted by $t_{\nu\mu}({\vec{k}_\parallel})$,
$\vec{k}_\parallel \mu$ and $\vec{k}_\parallel\nu$ labeling
flux-normalized states at the Fermi energy to the left and
right of the DW, respectively, including the spin labels.

A constant modulus for the local magnetization vector is
assumed; its direction may be represented by a single rotation
angle $\theta$ since we disregard the spin-orbit interaction.
$\theta$ varies along the $z$-direction but is constant in the $x,y$-directions. 
The exchange field of the DW can be diagonalized by a local gauge transformation
at the cost of an additional spin-rotation energy.  Instead of treating this term
by perturbation theory~\cite{Tatara2,Levy} we employ here the WKB approximation,
which has the important advantage of being valid also for vanishing exchange
splittings.

In order to understand the basic physics, let us first consider a simple
two-band model in which the plane
wave states with parallel wave vector
$\vec{k}_\parallel $ and energy $E_{\uparrow,\downarrow} =
\hbar ^2
k^2/2m \pm \Delta$ are modified by the domain wall. In the WKB method the spinor
wave functions are  multiplied by
$z$-dependent phase factors 
$\exp [ i\int^zdz'\sqrt{2mE_\pm(q(z'))/\hbar^2-k_\parallel^2}]$. The
eigenenergies of the local Hamiltonian in which
the gradient
$q(z)=\partial\theta/\partial z$ is taken to be constant are those of a ``spin
spiral''~\cite{Callaway}:
\begin{equation}
E_\pm (q)= \frac{\hbar ^2}{2m} \left[
k^2 + q^2/4 \pm \sqrt{ k_z^2 q^2 + p^4 } \right],
\label{eq:spsp}
\end{equation}
with $\Delta = \hbar ^2 p^2/2m$ and $k_z$ determined by $E_\pm (q)=E_F$.
The WKB-factor is imaginary for states propagating through
the whole DW and exponentially damped otherwise.  In our adiabatic approximation
we disregard all tunneling states, which is allowed in the limit
$\lambda_{DW}k_F \gg 1$.  
The eigenstates are
not pure spin-states: the DW/spin spiral system acts like a
spin-orbit scatterer to mix the two spin-directions. The DW conductance is thus
limited by the local band structure with the smallest number of modes at the
Fermi energy which is at the center of the DW where $q$ is maximal,
$q_{\rm max}= \pi /\lambda_{DW}$.
For perpendicular transport~\cite{density}:
\begin{equation}
G(q) = \frac{ e^2}{h} \frac{A}{2 \pi} \left\{
\begin{array}{lll}
\overline{k}_F^2 - q^2/4 & {\rm for} & q^2 \le 2 p^2 \\
\overline{k}_F^2 - p^2 + p^4/q^2 & {\rm for} & q^2 > 2 p^2 \\
\end{array}
\right. .
\label{eq:Gspsp}
\end{equation}
This equation holds when
$\overline{k}_F^2 > q^2/4 + p^2$, {\it i.e.} when both spin bands are occupied. 
Note that transport
parallel to the spin spiral is much less affected by the DW:
$G_{\parallel}(q) = G(0) + O(q^4)$.

In bulk transition metals in which $q^2 \ll p^2$, the DW-MR becomes
${\rm MR} = (G(q_{\rm max})-G(0))/G(0) = (\pi/2\lambda_{DW} \overline{k}_F)^2$  for
DW  independent of the exchange
splitting.   Using Eq.~(\ref{eq:Gspsp}), the Fermi wave vectors for one conduction
electron per atom, and the experimental width of the DW, we obtain the
numbers in Table 1 for Ni, Co and Fe. 
The effect appears to be very small and likely to be swamped by other
magnetoresistive effects such as the anisotropic or ordinary MR.  
The reason is clearly the smallness of the kinetic
spin rotation energy as compared to the exchange splitting, {\em i.e.}
$q^2 \ll p^2$. The DW only slightly deforms the Fermi spheres, resulting in
a tiny magnetoconductance. In transition metals,
however, many bands at the Fermi energy are much closer than
the exchange splitting. When spin-up and spin-down states close to
the Fermi energy are (nearly) degenerate, a DW which gives rise to 
a repulsive interaction between them may push the bands away from the Fermi
energy and reduce the conductance. Realistic band structures must be
used in order to evaluate the importance of these splittings.  To this end we 
carried out first-principles calculations of defect-free DWs in Ni(fcc),
Co(fcc), and Fe(bcc) within the local spin density approximation
(LSDA) to density functional theory (DFT). The open character of the leads
can be captured by the embedding Green function
technique\cite{Crampin,hoof1,hoof2} based on the
linearized-augmented plane wave method (LAPW) and the muffin-tin shape
approximation for the crystal potential. The transport coefficients and the
conductance of samples with arbitrary stacking of atomic monolayers with
non-collinear spins can be calculated with this method. The
technical details of the method are given in Refs.~\cite{hoof1,hoof2}.

In the adiabatic limit the DW
may be represented by a spin-spiral, which can be
computed using conventional band structure techniques by the generalized Bloch
theorem based on a combined translation and spin-rotation
operator~\cite{Sandratski}.  For narrow DWs a ``linear" model is more accurate,
in which we calculate the transmission (numerically) exactly for a magnetization
which is rotated by a constant rate $q_{\rm max}$ in a {\em finite} region of width
$\lambda_{DW}$ between single domain leads~\cite{hoof2}. 

The results are summarized in Table~I for the two
models just considered for
(i) experimental bulk DW widths and (ii) for DWs
of monolayer width, both for a total spin rotation $\pi$. Note the large
difference between the first-principles calculations and the two-band model.
Fig.~1 displays the width-dependent DW conductance as a function of the
magnetization rotation angle per monolayer for Fe and Ni, respectively. We
observe
a {\em linear dependence}, ${\rm MR} \sim - q_{\rm max}$, in clear contradiction 
of the two-band model (Eq.~(\ref{eq:Gspsp})).

We can understand these features using perturbation theory. The
spin spiral can be represented by an interaction Hamiltonian which contains two
operators $H_{\rm int}^{(1)} \sim q$ and $H_{\rm
int}^{(2)} \sim q^2$,
respectively~\cite{Tatara2}. The energy band structure of the bulk
ferromagnet and
thus the conductance is modified by this interaction. In non-degenerate
perturbation theory the first-order term corresponding to $H_{\rm
int}^{(1)}$ vanishes.
The second order term due to $H_{\rm int}^{(1)}$ and first order term due to
$H_{\rm int}^{(2)}$ both contribute to the order of $q^2$, which explains
the leading term in Eq.~(\ref{eq:Gspsp}). However, in the presence of
degeneracies simple perturbation theory breaks down. Instead, the
Hamiltonian must be diagonalized first in the subspace of (nearly)
degenerate states. The splitting of
the degenerate states is directly proportional to the matrix elements
of the interaction Hamiltonian, thus in leading order proportional to
$q$. As the energy splittings increase, conducting channels are removed 
from the Fermi surface and the conductance is reduced proportionally.
The linear dependence observed in Fig.~2 can thus be explained by the 
occurrence of many (nearly) spin-degenerate states close to the Fermi energy. 
Naturally, the MR is also much larger for closely spaced states which 
are not strictly degenerate. This explains the large
difference between the results for the two band model and the full band
structures in Table I.

We observe that the relative effect of the DW is still rather small,
smaller than the experiments by Gregg~\cite{Gregg} and smaller than the
theoretical
results by Levy and Zhang for very spin-asymmetric bulk
defect scattering~\cite{Levy}. Bloch DWs in thin films can be
significantly narrower than in bulk material~\cite{Kent}, which
means that the bulk DW magnetoresistance should be larger in thin films than the
bulk values in Table I, but still smaller than in~\cite{Gregg,Levy}. The present
calculations show unambiguously that the DWs {\em increase} the resistance. The
experimentally observed DW-induced decrease of the
resistance\cite{Otani,Rudiger} can therefore not be an intrinsic effect, 
but must be an as yet unidentified defect-related, size-related, or
other extrinsic phenomenon.   Previous results obtained by perturbation theory and a
two-band model~\cite{Tatara2,Levy} should be reconsidered in the light of the present
findings. Unfortunately, implementing degenerate perturbation theory for diffuse
systems~\cite{Tatara2} with realistic band structures appears to be
quite
cumbersome~\cite{Brataas}. 
In the recent work of Levy and Zhang \cite{Levy} 
the DW scattering is calculated on the basis of a
two-band model. In spite of the small forward spin-flip scattering in this model they can explain
a significant MR due to a strongly spin-dependent bulk defect scattering. The band structure
crowding at the Fermi surface enhances not only the backward scattering which causes the
ballistic MR discussed here, but also the forward scattering. The bulk defects might therefore be
less important than initially apparent. 

The DW scattering increases with $q=\partial\theta/\partial z$, which can be
achieved by reducing the DW width or by increasing the
winding number for a given width. Both operations become possible by
trapping a DW in a nanostructured ferromagnetic point contact.
Ballistic point contacts have been fabricated successfully in simple metals
\cite{Ralls,Holweg}, but not yet in ferromagnetic materials~\cite{Theeuwen}. 
When the magnetization on one side of the contact is pinned by shape anisotropy
or exchange biasing, the magnetization vector on the other side can be rotated
independently by rotating the sample in an external magnetic field.
The maximal effect is expected for an abrupt
domain wall, for which we predict a huge MR (see Table~I), much larger than
what has been be achieved with tunnel junctions of the same materials. The
material dependence on the angle between the two magnetizations (Fig.~2)
betrays again the importance of the details of the band structure. In a similar
fashion an $n \pi$-DW could be created by repeated rotation in the magnetic
field. The conductance is then predicted to decrease linearly with the number of
turns as in Fig. 1,
up to some value at which phase-slips occur, or the spiraling magnetization
spills out of the constriction. We stress that this somewhat naive picture needs to
be supported by micromagnetic calculations~\cite{Theeuwen}.

In conclusion, we presented and analyzed model and first-principles
calculations of
electron transport through magnetic domain walls. The large number of bands
close
to the Fermi surface causes a strong enhancement of the DW-MR as
compared to two-band calculations. Evidence that 
degeneracies at the Fermi surface of Fe, Co, and Ni can give rise to relatively
large effects is found. DWs always decrease the
ballistic conductance, causing a negative MR. The ballistic DW
magnetoresistance is found to be somewhat smaller than measured recently in
thin
films, which can be partly due to the reduction of domain wall widths in thin
films as compared to bulk ferromagnets. Trapping a domain
wall in nanostructured constrictions is predicted to give rise to a strongly
enhanced magnetoresistance.

We acknowledge helpful discussions with Jaap Caro,
Ramon van Gorkom, Junichiro Inoue, Andrew Kent, and Gen Tatara. This work is supported by the
"Stichting voor Fundamenteel Onderzoek der Materie" (FOM), and the "Nederlandse Organisatie
voor Wetenschappelijk Onderzoek" (NWO). We acknowledge benefits from the TMR Research Network
on "Interface
Magnetism" under contract No. FMRX-CT96-0089 (DG12-MIHT). G.E.W.B. would
like to
thank Seigo Tarucha and his group members for their hospitality at the NTT
Basic
Research Laboratories.

\begin{table}
\begin{center}
\begin{tabular}{lccc}
Property & Fe & Ni & Co \\ \hline
Crystal structure & bcc & fcc & fcc \\
Layer direction & (100) & (100) & (111) \\
$G_{sat}$ [$10^{15} \Omega^{-1} m^{-2}$] & 1.531 & 1.923 &
1.529 \\
DW thickness $\lambda_{DW}$ & & & \\
\hspace{0.5truecm} (in nm) & 40 & 100 & 15 \\
\hspace{0.5truecm} (in monolayers) & 276 & 570 & 72 \\
Spiral angle/monolayer  & $0.65^\circ$ & $0.32^\circ$ & $2.5^\circ$  \\
\\
DW-MR  & & & \\
\hspace{0.5truecm}  2-band model & -0.0008 \% & -0.0001 \% & -0.008 \% \\

\hspace{0.5truecm}  adiabatic model & -0.13 \% & -0.11\% &
-0.33\% \\
\hspace{0.5truecm} linear model & -0.39 \% & -0.16\% & -0.46\% \\
\hspace{0.5truecm} abrupt DW & -71 \% & -58 \% & -67 \% \\

\end{tabular}
\end{center}
\caption{Parameters for Fe, Ni and Co, calculated
saturation (single-domain) conductances and magnetoresistances (MR) as
defined in the
text. DW thicknesses are taken from Ref.~\protect\cite{Jiles}.  \label{tab:1}}
\end{table}

\begin{figure}[tbp]
\caption{Conductance of domain walls in Ni and Fe as a function of the magnetization
rotation
angle per monolayer, $\Delta \theta = \pi a/\lambda_{DW}$, where $a$ is the monolayer
width and $\lambda_{DW}$ the width of the domain wall.
Results are given for the adiabatic approximation (spin spiral) and the linear
approximation (see text). The bulk ballistic conductances are indicated by the
horizontal lines.}
\label{f:Ni}
\end{figure}

\begin{figure}[tbp]
\caption{Conductances of abrupt domain walls in Ni, Fe, and Co as a function of the
angle $\Delta \phi$ between the magnetization vectors of the bounding domains.}
\label{f:Angle}
\end{figure}


\begin{references}
\bibitem{density} For a slowly varying magnetization profile we expect that a
constant density approximation is more accurate than a constant chemical
potential
approximation. It can be shown that the correction to the conductance
due to the shift of the chemical potential is
$\delta G(q) / G(0) = - q^2 p^4 / 48 \overline{k}_F^6$ to the lowest order in
$\Delta /E_F$, which is negligible compared to $\Delta G(q) / G(0)$ in
Eq.~(\ref{eq:Gspsp}).
\end{references}
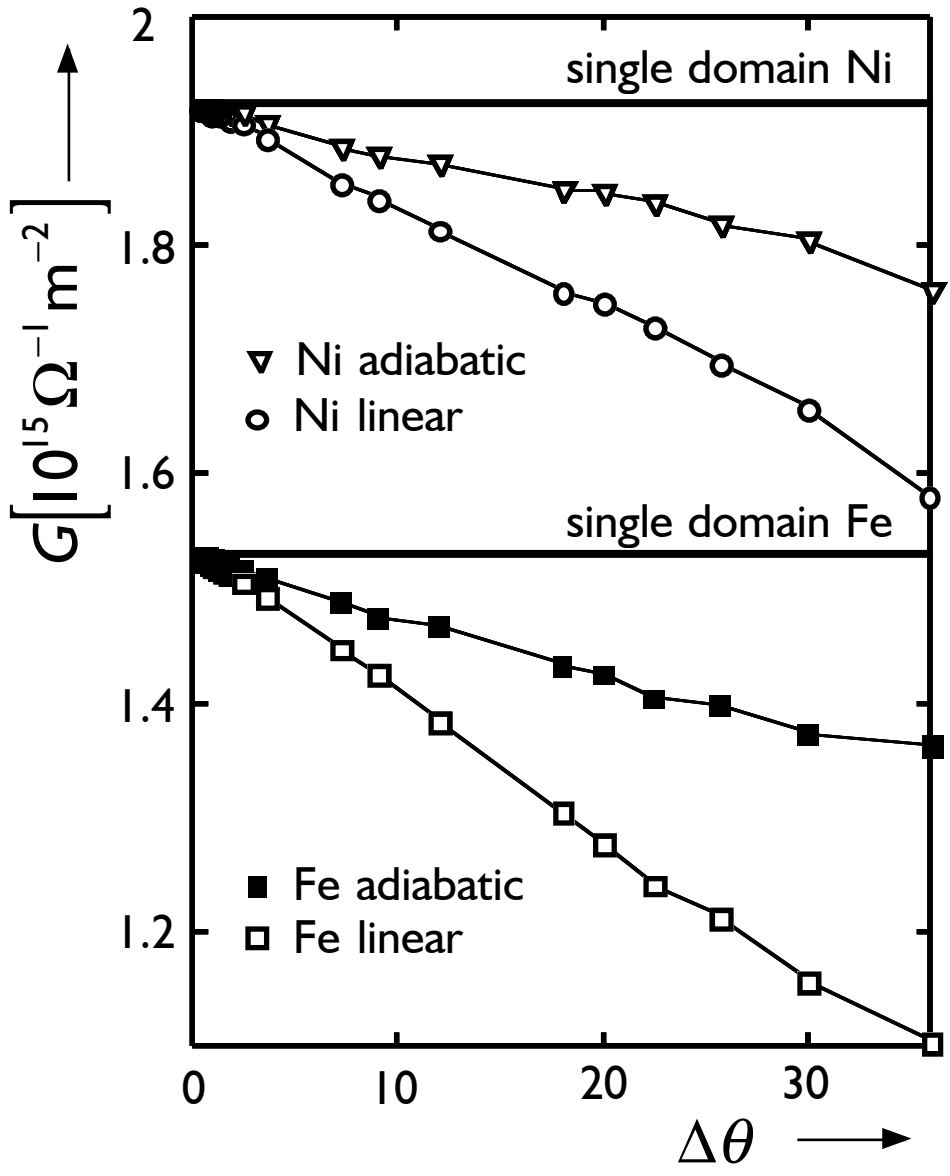
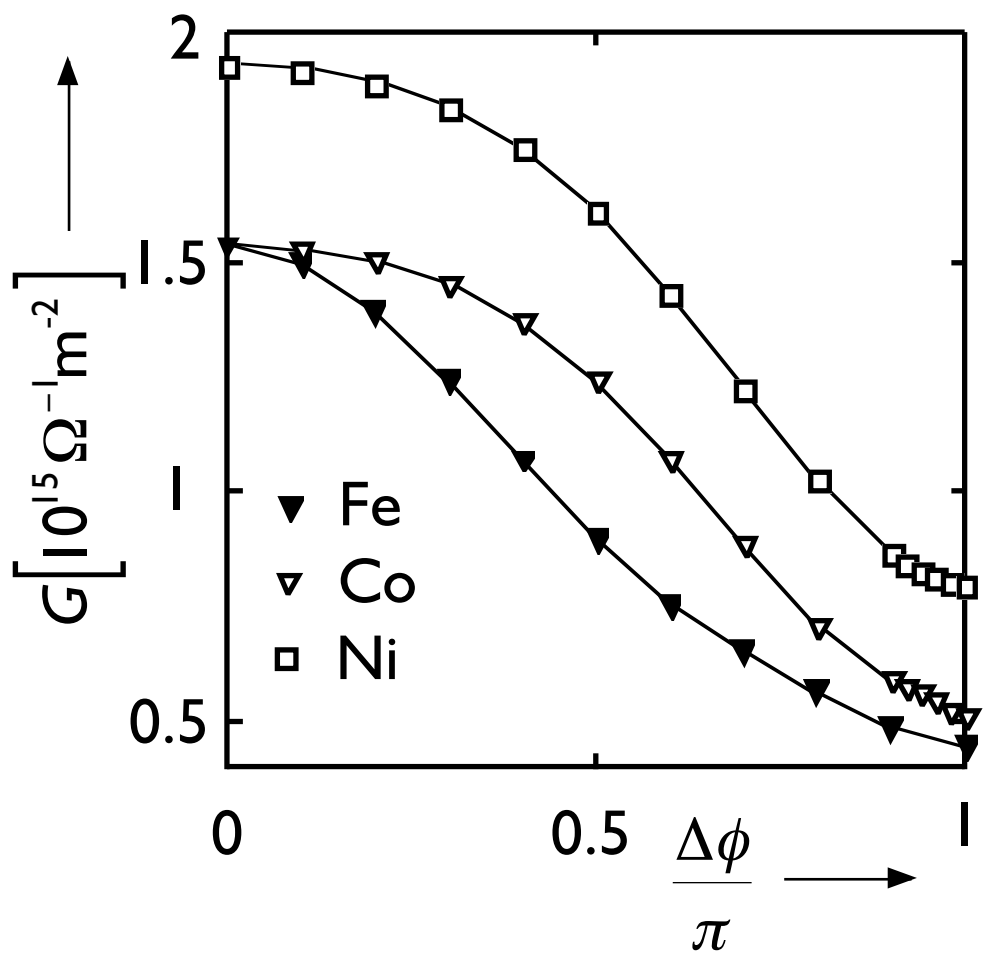
\end{document}